# Yet Another Pacman 3D Adventures

Serguei A. Mokhov

`mokhov@cse.concordia.ca`

Yingying She

`yy_she@cse.concordia.ca`

Concordia University, Montreal, Quebec, Canada

December 11, 2006

# Contents

# List of Figures

# Chapter 1

# Introduction

This game is meant to be extension of the overly-beaten pacman-style game (code-named "Yet Another Pacman 3D Adventures", or YAP3DAD) from the assignments and other projects with advance visual and gaming features, including game-in-a-game approach. The project is an open-source project published on `SourceForge.net`, specifically at `http://sf.net/projects/yap3dad`, for possible future development and extension.

## 1.1 Purpose

We attempted to study and employ a popular rendering engine, OGRE [The07], and re-implement the original game we had for the assignments of Pacman from the assignments and its original [r0d06]. That is quite an undertaking when learning the particularities of the new API, but the end result looks very promising. A few of our exploited targets of the engine are:

1. Visual

   - Illumination and fog.

   - Camera control. Provide different view point for Pacman and Ghosts.

   - Animate landscape rendering.

- Modularize model import with different meshes to be Ghosts and Pacman, etc.

- Animation. Applying keyframe with the OGRE API.

2. Backend

- Relative platform-independence: Linux and Windows, as even of the current version the games compiles and runs under both operating system flavors (tested on Fedora Core 4 and Windows XP Professional). It may work on other Operating Systems as well where `g++` or some similar compliant C++ compiler is available.

## 1.2   Beneficiaries

Open-Source, Educational, Gaming communities who are enthusiastic about Pacman-like games. We are trying to keep the development as a transparent process hoping it is useful for the folk like us, and we are considering releasing the game once it is more mature and stable to the OGRE community as a tutorial, while maintaining it as an entertaining tool and graphics features investigation platform "to try things out" and publish the results.

## 1.3   Challenges

- Different computer graphics techniques are tended to use in this project.

- Cross platform design.

- Multiplayer and network aspects.

- Game logic design.

### 1.3.1   Why is the problem hard?

The game is overly beaten and old and exists in too many variants to enumerate here. One of the challenges of making it playable and distinct. The technical challenges lay in the re-design and re-engineering of the code base and libraries, frameworks, networking, and multiplayer aspects. Each of them are hard on their own, and we cannot promise having them all done for this project, but having it open-source on `SourceForge.net` will allow further developers to either aid us or pick up the development effort in various areas.

### 1.3.2 What approaches have others tried?

Here we go over other existing open-source projects that attempt several approaches (but not necessarily complete the actual implementation) and their main features, which we may use as a source of ideas or features for our own implementation:

- "Pacman Arena [sub06] is a Pacman clone in full 3D with a few surprises. Rockets, bombs and explosions abound."

- "PANP [co06] is a new generation of the known game Pacman offering a 3D graphic interface, it provides 4 different views of game. The most outstanding feature is the possibility of playing net games against other players. It require OpenGL support."

- "A multiplayer team-based Pacman [bbc+06] game with exciting, fast-paced action."

- "Smart Pacman [ise06]: Take pacman's or ghosts' side in the challenge. Ghosts are controlled in the same way as units in an RTS game. You can play with AI or with a friend on a single keyboard."

- "Pacman Tag [dr06]: The playground game of "t" or "tag" combined with the arcade game "Pacman". One player is "it" and has to tag another player to stop their counter from ticking up. Whoever has been it for the least amount of time by the end of the game wins."

- "PacMan X [fm06] is a fully featured PacMan implementation."

- "Pacman for Sharp Zaurus [ben06] is a multi-board pacman game developed for the Qtopia environment. It was developed at Auburn University for a Senior Design project. The game is quite playable, with most features from the classic pacman game in place already."

- "Phoenix PacMan [sas06]: It is an advanced PacMan Game with four types of monsters, openable walls and lot of items. it written in Java, and requires JDK 1.3"

- "This project [dan06] intends to build a Winter version of Pac Man, which will be programmed under OpenGL/GTK environment."

- "Njam [mba06] is a pacman-like game with single/multiplayer/duel modes, networking support and integrated level editor. It runs on Linux, Windows, BeOS, OpenBSD, FreeBSD, MacOS X and MorphOS. It features great graphics, music, sound fx and a lot of levels to play."

- "Mango Quest [gp06] is a 3D arcade game using OpenGL and SDL, which aims to extend the pacman gameplay using a 1st person view and tons of special items. It runs on both Linux and Win32 platforms."

- "The game is PacMacro [lum06], the live action version of PacMan. We select a 5x6 block area of a major metropolitan area and play PacMan on a large scale. On the street we have the ghosts: Inky, Blinky, Pinky, and Clyde all chasing down the ever elusive PacMan."

- "A network multiplayer Pacman clone [mat06], tweaked for fast and exciting gameplay!"

- "A Multiplayer version of Pacman [kar06] with the focus on Deathmatch, programmed in the SDL. Later on it will feature Bots, A Leveleditor, more game types."

## 1.4 Tools and Techniques

We employed a variety of tools to design, develop, and maintain our project in Windows and Linux. The most prominent tool was a graphics rendering engine, OGRE, that presented a rather rich API to its implemented techniques for rendering and animation.

### 1.4.1 OGRE

OGRE (Object-Oriented Graphics Rendering Engine) [The07] is a scene-oriented, generally flexible 3D engine written in C++ designed to make it easier and more intuitive for developers to produce games and demos utilizing 3D hardware. The class library abstracts all the details of using the underlying system libraries like Direct3D and OpenGL and provides an interface based on world objects and other intuitive classes.

The features of OGRE include:

1. Object-Oriented Design, Plug-in Architecture, Save/Load System.

   - Simple, easy-to-use OO interface designed to minimize the effort required to render 3D scenes, and to be independent of 3D implementation i.e. Direct3D/OpenGL.

   - Flexible plugin architecture allows engine to be extended without recompilation.

   - Flexible plugin architecture allows engine to be extended without recompilation; Support for ZIP/PK3 for archiving of the media files.

2. Scripting

   - Scripted material language allows you to maintain material assets outside of your code (avoids unnecessary recompilation).

   - Scriptable multipass rendering.

3. Basic Physics, Collision Detection, Rigid Body

   - Controllers allow you to easily organize derived values between objects

   - Includes bindings for multiple 3rd party collision / physics systems (ODE, Novodex, and Tokamak)

4. Lighting, Per-vertex, Per-pixel, Lightmapping:

   - Can have an unlimited number of lights in the scene.

   - Supported through vertex and fragment programs.

5. Shadows, Shadow Mapping, Shadow Volume

   - Techniques supported: modulative and additive stencils, modulative projective.

   - Multiple stencil shadow optimizations.

   - Texture shadows fade out at far distance.

6. Texturing

   - Basic, Multi-texturing, Bumpmapping, Mipmapping, Volumetric.

   - Supports PNG, JPEG, TGA, BMP and DDS image file formats.

7. Supports High Level and Low Level Shaders, both Vertex and Pixel

   - Supports vertex and fragment programs (shaders), both low-level programs written in assembler, and high-level programs written in Cg or DirectX9 HLSL, and provides automatic support for many commonly bound constant parameters.

   - Supports GLSL

8. Scene Management, General, BSP, Octrees, Occlusion Culling, LOD

- Quite customizable, flexible scene management.
- Hierarchical scene graph and scene querying features.

9. Animation & Meshes

   - Inverse Kinematics, Skeletal Animation, Animation Blending
   - Mesh Loading, Skinning, Progressive

10. Surfaces & Curves

    - Splines, Biquadric Bezier patches for curved surfaces

11. Special Effects

    - Environment Mapping, Lens Flares, Billboarding, Particle System, Motion Blur, Sky, Water, Fog

12. Rendering, Fixed-function, Render-to-Texture, Fonts, GUI

    - Scriptable multipass rendering, Material LOD
    - Supports the complete range of fixed function operations
    - Support for multiple material techniques
    - Transparent objects automatically managed
    - Font system: TrueType fonts and precreated textures
    - 2D GUI System with Buttons, Lists, Edit boxes, Scrollbars, etc.

The website of OGRE is: `http://www.ogre3d.org/`.

### 1.4.2 Subversion

Subversion [Col07] is a version control tool for the source code and documentation repository. Similarly, to CVS [BddzzP+05], it keeps revisions of files and directories in the tree directory structure saving only modifications from the previous revision of a given file. It allows revert to any previous revision at any time and monitor the development. Our actual repository is hosted at `SourceForge.net`, specifically here, where one can browse it freely: `http://yap3dad.svn.sourceforge.net`. In Linux we used subversion command-line client `svn` to access the repository and in Windows we used `TortoiseSVN`.

### 1.4.3  GNU Make

GNU Make [SMSt06], or `gmake`, was a primarily used to manage the compilation of the project structure under Linux environment (with or without an IDE). A series of Makefiles are present throughout the code and documentation repository to compile the executable and even this LaTeX manual.

### 1.4.4  g++

`g++` is a C++ compiler from the GNU Compiler Collection [Vt05] used as a primary compiler to compile the source code under Linux from within Makefiles.

### 1.4.5  Microsoft Visual C++ 2003 .NET

This IDE was used in the lab on the Windows side to interface the development effort. Its project files are located in `src/scripts/win32/`.

### 1.4.6  KDevelop

This tool is an IDE in Linux to manage the project files, `autoconf`, and other settings on the Linux side. Most of its project files are located in `src/scripts/linux/`; other Makefiles are spread out through the actual source code, one per directory.

# Chapter 2

# Design and Implementation

This chapter debriefs on the approaches and methodology of the game (re)design and implementation.

## 2.1    Approach

As previously mentioned, we took the as a basis the original game from the assignments [r0d06], and then began adopting it to the OGRE semantics, so the features of the original game are not only preserved, but also enhanced, including compatibility of the map files. After rewriting the basics in OGRE, we managed to add a couple of things not present in the original.

## 2.2    Methodology

The original proposal included the steps below to re-design the game. Due to the time constrains of a course projects, only items in **bold** got looked into to some extends.

- **Re-engineering the existing code** and libraries **into frameworks so modules can be easily replaced if needed or new ones easily added**. This includes game logic engines, shading subsystem, networking multiplayer subsystem, **presentation subsystem**, **control** **subsystem**.

- Implementation/integration of the new lighting models through vertex and pixel shaders.

- **Porting all subsystems to the Linux/Windows**/MacOS as required. The implementation should never be one specific OS-centric to increase the user-base.

The YAP3DAD project followed the more or less traditional OO paradigm to re-engineer the original game. The summary of the classes produced as a result is in Figure 2.2. This also represents the architecture and model hierarchy of the project.

The classes like `PacExampleFrameListener` and `PacExampleApplication` follow a typical application design that most of the OGRE applications do. They define initialization of the camera, response to the input events and time events (which are further used in the key-frame animation).

`CMazeEntity` represents just about any entity in the maze map – walls, pellets, power pills, ghosts, and the pacmen, and other entities alike that can be fit onto a typical grid. It has position, abstract API of `draw()`, `animate()`, and the like, which the derivatives override as the need. A `CMaze` consists of a collection of cells, depicted by walls (`CWall`, zero to up to 4), a floor tile `CFloor`. Then, there are pellets, and their extension – power pills, that are parallel in the hierarchy to the maze cells. The reason for this is the use of static geometry for the stationary scene elements to improve performance (static geometry is preprocessed once and then bunched up to be fed to the GPU in batches efficiently). Animated pieces cannot be grouped along with static geometry, so they require a separate hierarchy of scene nodes. Next, there are ghosts and pacmen. There could be more than one of each, (even pacman) the way the game was redesigned, but the actual multiplayer game play was not implemented at the time of this writing.

The `CGameEnv` came from the original game [r0d06], which is primarily responsible for loading and parsing the map files because we kept the same

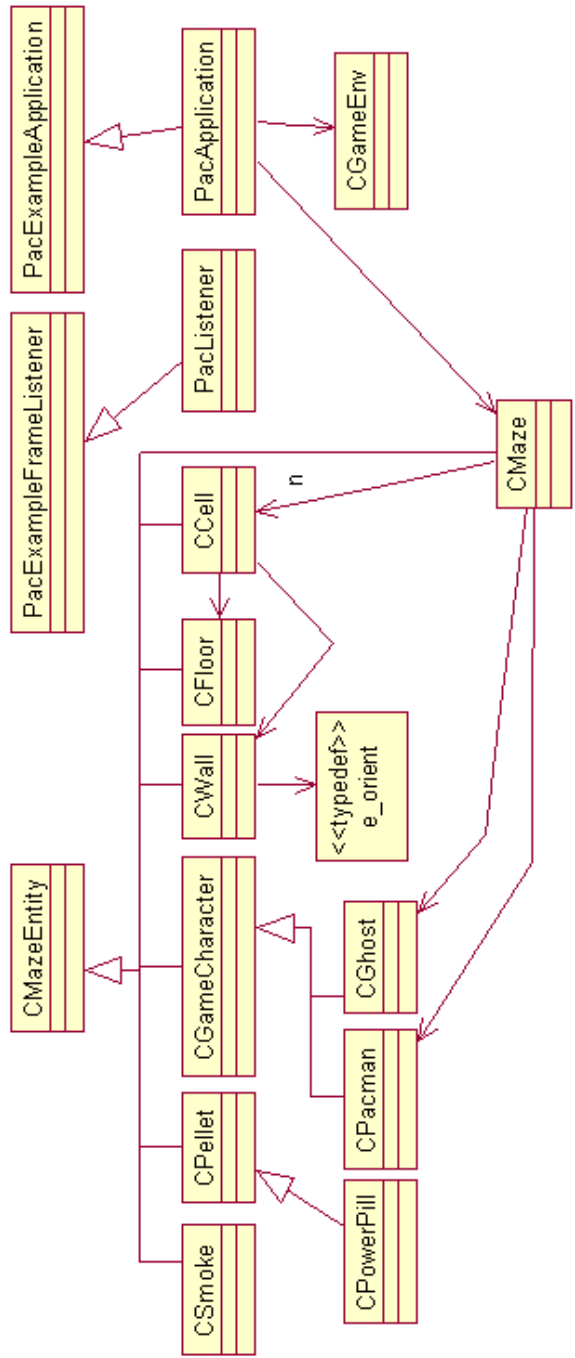

Figure 2.1: YAP3DAD High-Level Class Diagram.

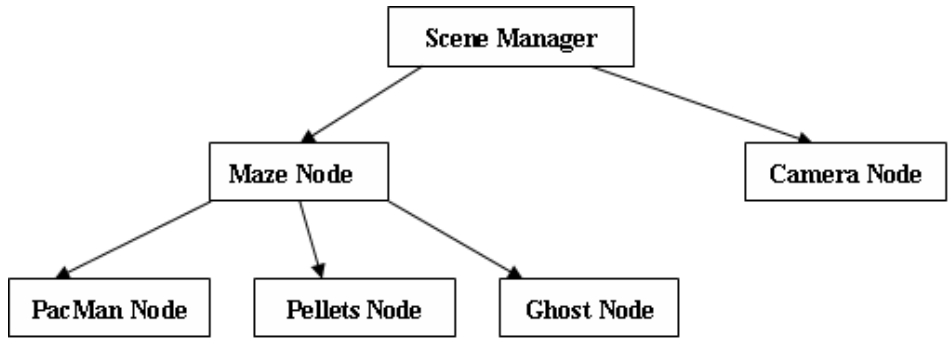

Figure 2.2: OGRE Scene Division Hierarchy

map file format, so any maps developed for the original game would work in our variant.

## 2.2.1 OGRE Tools

A typical OGRE program is structured using an `SceneManager`, `Entity`, and `SceneNode` basis system classes. Everything in the scene is control by the `SceneManager` (it is possible to have multiple scene managers that manipulate the scene nodes depending on some optimization techniques depending on what you are trying to render as a terrain, environment, etc, e.g. `BSPSceneManager`). Every movable object in the scene is an `Entity`. Every object in the scene is a `SceneNode` placed in the scene hierarchy (tree). Our own class hierarchy presented above followed more or less similar structure when wrapping around OGRE primitives.

In addition, this YAP3DAD uses scripting based texture, particles, 2D menu rendering system from OGRE. For animation, the project uses mesh based skeleton animation supported by OGRE.

## 2.2.2 Game Logic

One of a particularly hard issues was to adapt the original game logic to the new mindset of events and API, after learning it. A fun element is also important here when trying out new things, and so it's not too boring while playing or even testing.

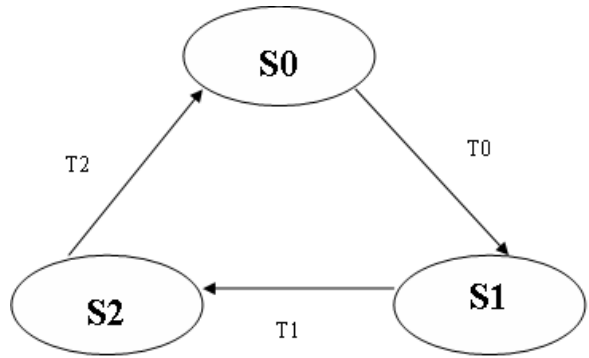

Figure 2.3: States of a Robot Ghost

The game logic of original Pacman game is straightforward and simple. In order to make it more interesting, game logic design is another important design part we concentrated on this project.

YAP3DAD uses a **Finite State Machine** in here. For example, robot ghosts have three states and three transitions in the game, as in Figure 2.2.2:

- **S0**: Robot Ghost is walking randomly in Maze

- **S1**: Robot Ghost meets the Pacman

- **S2**: Robot Ghost dies

- **T0**: Robot Ghost transitions from S0 to S1

- **T1**: Robot Ghost transitions from S1 to S2

- **T2**: Reset Robot Ghost to S0

At the same time, we display the game art as appropriate. YAP3DAD imported **Ninja, Robot**, and **Fish** meshes as **Pacman**, **Power Pills and Ghosts**, and **Pellets** due to their readily availability in OGRE. It would be easy enough to load other models, and the only changes required would be to point to the class which meshes and states to use as the defaults. In addition, we use different named animations (e.g. 'Walk', 'Run', 'Jump', 'Shoot', etc.) encoded with the supplied model and its skeleton to represent different statuses of the **Pacman** and **Pellets**. Theses features increase the attraction of the game.

### 2.2.3 Modular Design

Due to the modular design of the proposed frameworks, the missing implementations can be replaced by stubs or local versions until they are replaced by the real version when time permits. The design will is done in such a way that the crucial steps will not affect the overall project status, and will be "rolled in" in the subsequent builds.

# Chapter 3

# Mini User Manual

This chapter details some aspects of running and compiling YAP3DAD from sources.

## 3.1 Prerequisites

If you want to compile YAP3DAD you need to have some tools. If you download this project from `http://sourceforge.net/projects/yap3dad/` Please follow the steps in Section 3.1 and Section 3.2 if you want to compile the source code. If you already have the executable, please refer to the Section 3.3.

**Windows XP**

1. Your system should have Visual C++ .NET 2003 (7.1)

2. Install OGRE 1.2.4 SDK for Visual C++ .NET 2003 (7.1) in your computer.

   - For example, you can install the OGRE SDK in `C:\OgreSDK`
   - You can download it from OGRE website: `http://www.ogre3d.org/index.php?option=com_remository&Itemid=57&func=fileinfo&filecatid=41&parent=category`

**Linux (Fedora Core 4, but others should also work)**

1. Your system should have **g++** installed, a part of GCC.

2. Install OGRE 1.2.4 SDK for Linux by compiling it through the typical procedure of `./configure`, `make`, `make install`. You would need the Cg and DevIL libraries installed.

   You can download it from OGRE website. *http://www.ogre3d. org/index.php?option=com_remository&Itemid=57&func=fileinfo&filecatid= 41&parent=category*

## 3.2 Compilation

### 3.2.1 Windows XP

1. Before build solution for the project in VC++ 2003, please copy *Pac-Pacman.overlay* and *Pac-PacPoint.overlay* from

   `{projectfolder}\media\overlay`

   to

   `c:\OgreSDK\media\overlay`

2. Before build solution for the project in VC++ 2003, please copy the whole folder

   `{projectfolder}\map`

   to

   `C:\OgreSDK\bin`

3. Build the solution for the project Visual C++ .Net 2003 (7.1) (located in **src/scripts/win32**).

4. After build the solution, the project will create a executable file **yap3dad.exe** in `C:\OgreSDK\bin\debug`. Double click this file, you can run it in your computer.

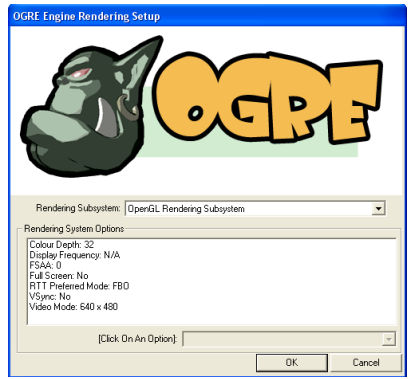

Figure 3.1: Config panel in Windows

### 3.2.2 Linux

1. First, copy all `*.cfg` files from `src/scripts/linux` to just `src`.

2. then, to compile YAP3DAD, `cd` to the directory `src` and type `make -f Makefile.in`,

3. then to run it, just type `./yap3dad`

## 3.3 User Manual

1. After you run **_yap3dad_**, there will be a OGRE configuration window pop up.

   - If you have the zip version of YAP3DAD, after unzip the project, you will find this file in `projectfolder\bin\debug`
   - If you follow the step in 3.1 and 3.2, this file is in `C:\OgreSDK\bin\debug`.

   The configuration window is as below (first as in Windows and then as in Linux):

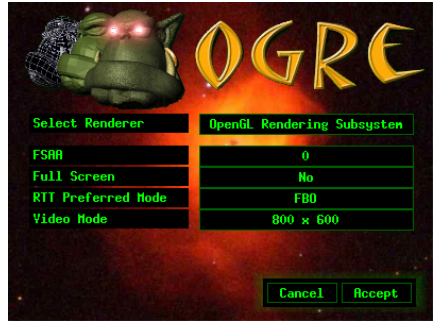

Figure 3.2: Config panel in Linux

You can choose **Rendering Subsystem** and **Rendering System Options** in this window (under Linux only OpenGL would be available). And click **OK** to start the project.

2. User Manual Panel – Scene Control

- If you press **H**, the User Manual Panel will show on the top left side of the rendering window.

- **Left Arrow**: Rotate camera counter-Clockwise

- **Right Arrow**: Rotate camera clockwise

- **W or Up**: Forward

- **S or Down**: Backward

- **A**: Step left

- **D**: Step right

- **PgUp**: Move upwards

- **PgDown**: Move downwards

- **F**: Toggle frame rate stats on/off

- **R**: Render mode

- **T**: Cycle texture filtering as bilinear, trilinear, or anisotropic

- **P**: Toggle on/off display of camera position / orientation

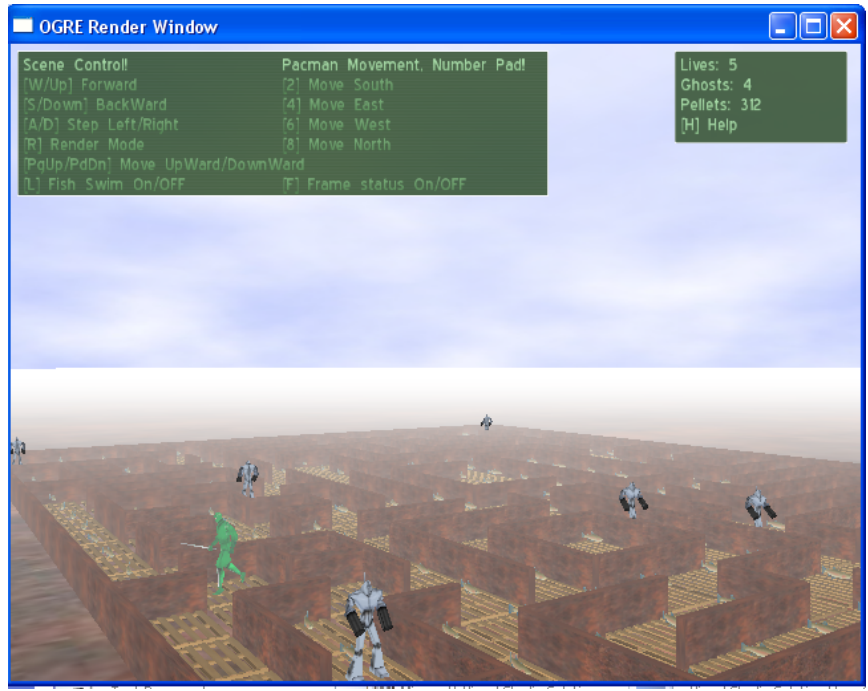

Figure 3.3: Help Overlay

- **L**: Fish swim turn on/off

3. User Manual Panel– Pacman Control

   The control over Pacman moving along the maze is using the [2], [4], [6], [8] keys in number pad.

   - **2**: Pacman goes south
   - **4**: Pacman goes west
   - **6**: Pacman goes east
   - **8**: Pacman goes North

4. FPS Info Panel

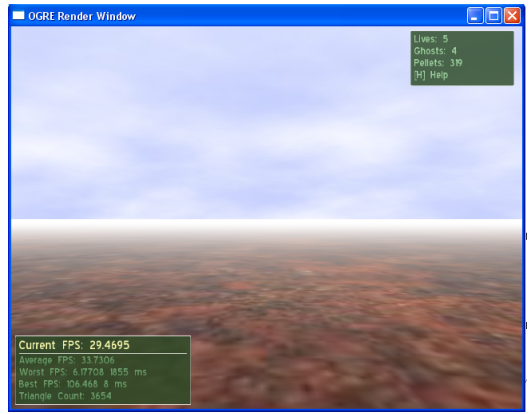

Figure 3.4: FPS (and other stats) Overlay

If you press **F**, the FPS panel will show on the down left side of the rendering window. The FPS panel show information about *Current FPS*, *Average FPS*, *Worst FPS*, *Best FPS*, *Triangle Counts*.

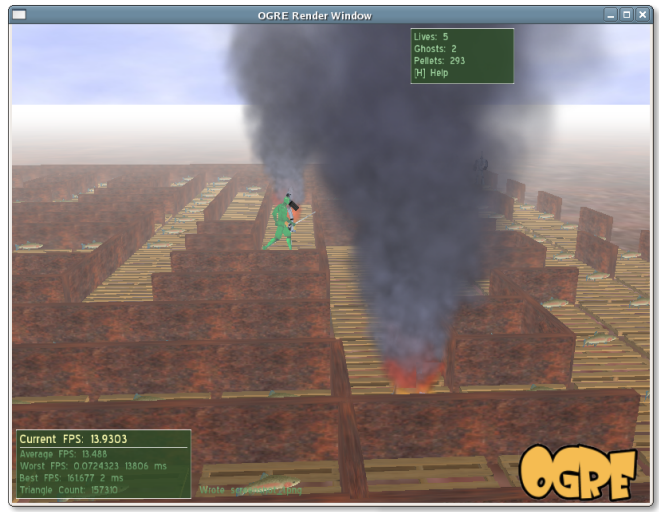

Figure 3.5: FPS (and other stats) Overlay in Game Action of Burning Ghosts (Linux)

# Chapter 4

# Future Work

This chapter summarizes the TODO/Wishlist items for the next few iterations of the project.

## 4.1   TODO Items

1. Visual

   - Pixel and vertex shaders to achieve realistic lighting effects (other than the Phong lighting).
   - Landscape rendering. Instead of drawing walls based on maps, drawing buidings, trees etc.
   - More sophisticated collision detection.
   - Vertical movement.
   - First-person action.
   - Slow-motion ghost-busting.

2. Backend

   - **Frameworked inner games (easy to add inner games), i.e. games-in-a-game aspect.** Games-in-a-game makes this game have more entertaining attractions. If designed in a pluggable framewored way, the mini-games can be interchanged, swapped-in, or swapped-out at any point in time in the development.

- **Multiplayer concept over network.** Multiplayer online games are the most popular video games in game industry. Human-to-huamn model makes video game more attractive than human-to-computer model.

### 4.1.1 Why

To our knowledge no-one has tried advanced GPU programming with custom lighting models through vector and pixel shaders in a similar style pacman-like game.

Additionally, for this game people didn't seem to try the game-in-a-game approach is for entering special rooms or in between the levels there can be other related mini-arcade or strategy, or logic games embedded in.

### 4.1.2 What

- Implementation/integration of the new lighting models through vertex and pixel shaders.

- Implementation/integration of the multiplayer concept over the network.

- Implementation/integration games-in-a-game levels.

Some of the aspects, e.g. networking, can be ported from a similar project, previously listed.

# Chapter 5

# Conclusion

Most modern computer games can be split into three parts: the game engine, the game logic, and the game art. In this game project, we develop a game based on 3D rendering engine OGRE. OGRE is an open-source game engine. It has its own programming structure which is different and abstracted from OpenGL and DirectX. We got the feel of OGRE and believe the game has a high potential to move forward using OGRE's features for quick prototyping to the release-quality game.

The game is an open-ended project. We only know when we are done is when we reach a certain milestone, which would be followed by the next one, and the next one. Thus, we will try to accomplish goals that we mentioned in the Future Work in the subsequent builds. Having it as an open-source project publicly available increases chances of a longer lifespan and future work to be done. Therefore, we will simply continue this project as open source project and keep adding new features in our free time.

We are very excited to work in further, and should *you* have any interest in, you are welcome to join the project! Just drop us a line:

- Serguei, `mokhov@encs.concordia.ca`
- Ying Ying, `yy_she@encs.concordia.ca`
- Project Mailing List, `yap3dad-devel@users.sourceforge.net`
- Project Page, `http://sourceforge.net/projects/yap3dad`